\documentclass[twocolumn,showpacs,preprintnumbers,showkeys,superscriptaddress]{revtex4}
\usepackage{amssymb}
\usepackage{graphicx}
\usepackage{dcolumn}
\usepackage{bm}
\usepackage{appendix}

\def\eqref#1{Eq.~(\ref{eq:#1})}

\begin{document}

\title{Accuracy of the new pairing theory and its improvement}
\author{L. Y. Jia}  
\affiliation{Department of Physics, University of Shanghai for
Science and Technology, Shanghai 200093, P. R. China}
\affiliation{Department of Physics, Hebei Normal University,
Shijiazhuang, Hebei 050024, P. R. China}

\date{\today}

\begin{abstract}

Recently I proposed \cite{Jia_4} a new method for solving the
pairing Hamiltonian with the pair-condensate wavefunction ansatz
based on the Heisenberg equations of motion for the density matrix
operators. In this work an improved version is given by deriving the
relevant equations more carefully. I evaluate both versions in a
large ensemble with random interactions, and the accuracy of the
methods is given statistically in terms of root-mean-square
derivations from the exact results. The widely used variational
calculation is also done and the results and computing-time costs
are compared.

\end{abstract}


\vspace{0.4in}

\maketitle

\section{Introduction}

Pairing correlation has long been recognized in nuclei
\cite{BCS_nucl1} and influences all the properties of the latter,
such as mass, gap of excitation energy, and moment of inertia
\cite{BCS_nucl2}. In general, any mean-field treatment of the nuclei
needs to account it somehow to get reasonable results. Among the
methods \cite{BCS_book} the most popular one may be the
Bardeen-Cooper-Schrieffer (BCS) theory \cite{BCS}, or its advanced
version the Hartree-Fock-Bogoliubov (HBF) theory \cite{BCS_nucl2},
where pairing correlation is considered by introducing
quasi-particles and writing the ground state as a vacuum of the
latter. But there are disadvantages of breaking the exact particle
number and need for a unphysical minimum pairing strength
\cite{Bohr_book, Ring_book}.

A common improvement is to use the pair-condensate [Eq. (\ref{gs}),
the BCS wavefunction projected onto good particle number] as the
ground state wavefunction \cite{Dietrich}. Usually the criteria to
determine the variational parameters is minimizing the energy in the
variation principle (variation after projection BCS) \cite{Dietrich,
Robledo, Flocard, Braun, Dukelsky, Delft}. In Ref. \cite{Jia_4} I
proposed a new criteria based on the Heisenberg equations of motion
(EOM) for density matrix operators, as the lowest-order (mean-field)
results of the generalized density matrix (GDM) formalism
\cite{kerman63,BZ,Zele_ptps,shtokman75,Jia_1,Jia_2,Jia_3}. The
method was applied to the calcium isotopes with the well known FPD6
interaction \cite{fpd6}.

In this work I give an improved version of the method. In Ref.
\cite{Jia_4} the relevant equations are derived assuming that
neighboring even-even nuclei have the same Hartree-Fock (HF)
single-particle energies and occupation numbers; here I do it more
carefully allowing the latters to be different. To see the validity
of the approaches, both versions are applied to a large ensemble
with $1000$ examples and random interactions; the results are good
in almost all the examples. I also compare the results and necessary
computing-time costs with those of the variation principle. In Sec.
\ref{Sec_form} I derive the improved version of the GDM pairing
theory. Then in Sec. \ref{Sec_rand_ens} the approaches are evaluated
in the large random ensemble. Finally Sec. \ref{Sec_sum} summarizes
the work and discusses further directions.

\section{Formalism  \label{Sec_form}}

The pairing theory in Ref. \cite{Jia_4} was derived as the
lowest-order (mean-field) results of the complete GDM formalism. In
fact the derivation could be simpler and more clear if we focus on
the mean fields and do not introduce collective (quadrupole)
phonons. Also, in Ref. \cite{Jia_4} I made the assumption that
neighboring even-even nuclei have the same HF single-particle
energies and occupation numbers. Below I derive an improved version
of the pairing theory abandoning the above assumption.

As before, the ground state of the $2N$-particle system is assumed
to be an $N$-pair condensate,
\begin{eqnarray}
|\phi_N\rangle = \frac{1}{\sqrt{\chi_{N}}} (P^\dagger)^{N} |0\rangle
, \label{gs}
\end{eqnarray}
where $\chi_{N}$ is the normalization factor, and $P^\dagger$ is the
pair creation operator
\begin{eqnarray}
P^\dagger = \frac{1}{2} \sum_1 v_1 a_1^\dagger a_{\tilde{1}}^\dagger
. \label{P_dag}
\end{eqnarray}
In Eq. (\ref{P_dag}) the summation runs over the entire
single-particle space. $|\tilde{1}\rangle$ is the time-reversed
level of the single-particle level $|1\rangle$. The pair structure
$v_1$ are parameters to be determined by the theory.

With the antisymmetrized fermionic Hamiltonian
\begin{eqnarray}
H = \sum_{12} \epsilon_{12} a_1^\dagger a_2 + \frac{1}{4}
\sum_{1234} V_{1234} a_1^\dagger a_2^\dagger a_3 a_4 ,  \label{H_f}
\end{eqnarray}
I calculate the exact Heisenberg equations of motion for the density
matrix operators $R_{12} \equiv a_2^\dagger a_1$ and $K_{12} \equiv
a_2 a_1$,
\begin{eqnarray}
[R_{12}, H] = [\epsilon , R]_{12}  \nonumber \\
- \frac{1}{2} \sum_{345} V_{5432} a_5^\dagger a_4^\dagger a_3 a_1 +
\frac{1}{2} \sum_{345} V_{1345}
a_2^\dagger a_3^\dagger a_4 a_5 ,  \label{eom_R} \\
{[}K_{12}, H] = (\epsilon K)_{12} - (\epsilon K)_{21} + \frac{1}{2}
\sum_{34} V_{1234} K_{43} \nonumber
\\
- \frac{1}{2} \sum_{345} V_{1543} a_5^\dagger a_4 a_3 a_2 +
\frac{1}{2} \sum_{345} V_{2543} a_5^\dagger a_4 a_3 a_1 .
\label{eom_K}
\end{eqnarray}
Terms on the right-hand sides of Eqs. (\ref{eom_R}) and
(\ref{eom_K}) are read as matrix multiplications; for example,
$[\epsilon, R]_{12} = (\epsilon R)_{12} - (R \epsilon)_{12} = \sum_3
\epsilon_{13} R_{32} - \sum_3 R_{13} \epsilon_{32}$.

On the pair condensate (\ref{gs}), the density matrices are
``diagonal'':
\begin{eqnarray}
\rho^N_{12} \equiv \langle \phi_{N} | a_2^\dagger a_1 | \phi_{N}
\rangle = \delta_{12} n^N_1 ,  \label{rho_def}  \\
\kappa^N_{12} \equiv \langle \phi_{N-1} | a_2 a_1 | \phi_{N} \rangle
= \delta_{\tilde{1}2} s^N_1 .  \label{kappa_def}
\end{eqnarray}
In practical shell-model calculations usually each single-particle
level has distinct spin and parity, thus the mean fields are also
``diagonal'':
\begin{eqnarray}
f^N_{12} \equiv \epsilon_{12} + \sum_{34} V_{1432} \rho^N_{34} =
\delta_{12} e^N_1 ,  \label{f_def}  \\
\delta^N_{12} \equiv \frac{1}{2} \sum_{34} V_{1234} \kappa^N_{43} =
\delta_{1\tilde{2}} g^N_1 .  \label{delta_def}
\end{eqnarray}
The pairing mean field $\delta^N$ should not be mixed with the
Kronecker delta $\delta$.

Now I take matrix elements of the exact equations of motion between
the pairing ground states (\ref{gs}): equation (\ref{eom_R}) between
those of the same nuclei (``$\langle \phi_N |$'' and ``$| \phi_N
\rangle$''); equation (\ref{eom_K}) between those of neighboring
even-even nuclei (``$\langle \phi_{N-1} |$'' and ``$| \phi_N
\rangle$''). On the left-hand side of Eq. (\ref{eom_K}) we have $
\langle \phi_{N-1} | [K_{12}, H] | \phi_N \rangle = \langle
\phi_{N-1} | K_{12} H | \phi_N \rangle - \langle \phi_{N-1} | H
K_{12} | \phi_N \rangle \approx (E_N - E_{N-1}) \kappa^N_{12}$,
where $E_N$ and $E_{N-1}$ are the ground state energies,
$H|\phi_N\rangle \approx E_N|\phi_N\rangle$ and $H|\phi_{N-1}\rangle
\approx E_{N-1}|\phi_{N-1}\rangle$. Similarly for Eq. (\ref{eom_R})
we have $\langle \phi_N | [R_{12}, H] | \phi_N \rangle \approx (E_N
- E_N) \rho^N_{12} = 0$.  On the right-hand sides the ``two-body
density matrices'' are approximated in the following way:
\begin{eqnarray}
\langle \phi_N | a_4^\dagger a_3^\dagger a_2 a_1 | \phi_N \rangle
\approx \langle \phi_N | a_4^\dagger a_1 | \phi_N \rangle
\langle \phi_N | a_3^\dagger a_2 | \phi_N \rangle  \nonumber \\
- \langle \phi_N | a_4^\dagger a_2 | \phi_N \rangle \langle \phi_N |
a_3^\dagger a_1 | \phi_N \rangle     \nonumber \\
+ \langle \phi_N |
a_4^\dagger a_3^\dagger | \phi_{N-1} \rangle \langle \phi_{N-1} |
a_2 a_1 |
\phi_N \rangle ,   \nonumber \\
= \rho^N_{14} \rho^N_{23} - \rho^N_{24} \rho^N_{13} + \kappa^{N
\dagger}_{34} \kappa^N_{12} ,
\label{fac1} \\
\langle \phi_{N-1} | a_4^\dagger a_3 a_2 a_1 | \phi_N \rangle
\nonumber \\
\approx \langle \phi_{N-1} | a_4^\dagger a_1 |
\phi_{N-1} \rangle
\langle \phi_{N-1} | a_3 a_2 | \phi_N \rangle  \nonumber \\
- \langle \phi_{N-1} | a_4^\dagger a_2 | \phi_{N-1} \rangle \langle
\phi_{N-1} | a_3 a_1 | \phi_N \rangle     \nonumber \\
+ \langle
\phi_{N-1} | a_4^\dagger a_3 | \phi_{N-1} \rangle \langle \phi_{N-1}
| a_2
a_1 | \phi_N \rangle  \nonumber \\
= \rho^{N-1}_{14} \kappa^N_{23} - \rho^{N-1}_{24} \kappa^N_{13} +
\rho^{N-1}_{34} \kappa^N_{12} .  \label{fac2}
\end{eqnarray}
Equations (\ref{fac1}) and (\ref{fac2}) would be exact if the ground
states were taken as single-particle Slater determinants: the
right-hand sides were just the fully contracted terms in Wick's
theorem. Here they are approximations because the ground states are
taken as pair condensates (\ref{gs}). Finally Eqs. (\ref{eom_R}) and
(\ref{eom_K}) become
\begin{eqnarray}
0 = [f_N, \rho_N] - \kappa_N \delta_N^\dagger + \delta_N
\kappa_N^\dagger ,  \label{eom_r}   \\
(E_N - E_{N-1}) \kappa_N = f_{N-1} \kappa_N + \kappa_N f_{N-1}^T  \nonumber \\
+ \delta_N - \delta_N \rho_{N-1}^T - \rho_{N-1} \delta_N
. \label{eom_k}
\end{eqnarray}
Under the ``diagonal'' properties (\ref{rho_def}),
(\ref{kappa_def}), (\ref{f_def}), and (\ref{delta_def}), Eq.
(\ref{eom_r}) is satisfied automatically, and Eq. (\ref{eom_k})
becomes
\begin{eqnarray}
E_N - E_{N-1} = 2 e^{N-1}_1 + g^{N}_1 ~ \frac{2 n^{N-1}_1 -
1}{s^{N}_1} .  \label{eq_final2}
\end{eqnarray}
Equation (\ref{eq_final2}) is the main equation of the improved
theory. It implies that the right-hand side is independent of the
single-particle label $1$, by which the parameters $v_1$ in Eq. (2)
are fixed. The main equation of the old theory, Eq. (19) in Ref.
\cite{Jia_4}, corresponds to replacing $e^{N-1}_1$ and $n^{N-1}_1$
by $e^{N}_1$ and $n^{N}_1$. Equation (\ref{eq_final2}) includes the
well-known particle-particle random phase approximation
\cite{Ring_book} as its special case of $N=1$ ($n^0 = 0$, $e^0 =
\epsilon$). The normalization factor $\chi_N$ (\ref{gs}), occupation
numbers $n^N_1$ (\ref{rho_def}), pair-transfer amplitudes $s^N_1$
(\ref{kappa_def}), mean fields $e^N_1$ (\ref{f_def}) and $g^N_1$
(\ref{delta_def}) are functions of the pair structures $v_1$
(\ref{P_dag}); their functional forms, as the ``kinematics'' of the
system, have already been given in Eqs. (23) and (24) of Ref.
\cite{Jia_4} and are not repeated here.

In the next section I take the pairing Hamiltonian:
\begin{eqnarray}
\epsilon_{12} = \delta_{12} \epsilon_1, ~ V_{1234} = -
\delta_{2\tilde{1}} \delta_{3\tilde{4}} G_{13} \label{H_pairing}
\end{eqnarray}
in Eq. (\ref{H_f}). Consequently the mean fields (\ref{f_def}) and
(\ref{delta_def}) become
\begin{eqnarray}
e^N_1 = \epsilon_1 - G_{11} n^N_1 ,~~~ g^N_1 = \frac{1}{2} \sum_2
G_{12} s^N_2 .     \label{eg_pair}
\end{eqnarray}
I apply both versions of the theory to a large random ensemble; for
convenience I call Eq. (\ref{eq_final2}) the ``GDM2'' pairing
theory, and Eq. (19) in Ref. \cite{Jia_4} the ``GDM1'' pairing
theory.

\section{Random Ensemble  \label{Sec_rand_ens}}

The ``GDM1'' theory was applied in Ref. \cite{Jia_4} to calcium
isotopes with the FPD6 interaction and the results are good. Here I
would like to consider both the ``GDM1'' and ``GDM2'' theories in a
large ensemble with random interactions; consequently we can speak
statistically the accuracy of the theories in terms of the
root-mean-square derivations from the exact results. The variational
calculation with the trial wavefunction (\ref{gs}) is also performed
and the accuracy and computing-time cost are compared.

The random ensemble has $1000$ examples with different parameters
determined in the following way. For each example, I first pick up
the single-particle levels randomly from the pool $2j = 1, 3, 5, 7,
9, 11, 13$. Each angular momentum $j$ has a $40 \%$ probability of
being selected; the selected ones (at least two) constitute the
model space. Second, the single-particle energies $\epsilon_1$
(\ref{H_pairing}) are determined randomly following the uniform
distribution from $- 10$ MeV to $0$. Third, I pick up the ``pairing
strength'' $G_{\rm{max}}$ as a random number following the uniform
distribution from $0$ to $2$ MeV, then the pairing matrix elements
$G_{12}$ (\ref{H_pairing}) are distributed uniformly from $0$ to
$G_{\rm{max}}$. Finally, the number of pairs $N$ is determined
following the uniform distribution from $1$ to $\Omega - 1$, where
$2\Omega = \sum_j (2j+1)$ is the maximal particle number allowed by
the model space.

I perform four sets of calculations for the ensemble: two GDM
calculations ``GDM1'' and ``GDM2'', variational calculation ``VAR'',
and the exact calculation. (In this work the exact calculation is
done by diagonalization in spaces with fixed seniority \cite{Sen1,
Sen2}. It can also be achieved by the Monte Carlo algorithm
\cite{MC1, MC2, MC3}; or the Richardson's method in some special
cases \cite{Richardson, Richardson_rmp, Richardson_new}.) For the
variational calculation of the pair-transfer amplitudes $s_1 =
\langle \phi_{N-1} | a_{\tilde{1}} a_1 | \phi_N \rangle$
(\ref{kappa_def}), I show two sets of results. In ``VAR1'' the pair
structure $v_1$ (\ref{P_dag}) in $|\phi_{N-1}\rangle$ and
$|\phi_{N}\rangle$ are the same, given by minimizing
$\langle\phi_{N}|H|\phi_{N}\rangle$; while in ``VAR2'' $v_1$ in
$|\phi_{N-1}\rangle$ and $|\phi_{N}\rangle$ are different, given by
minimizing $\langle\phi_{N-1}|H|\phi_{N-1}\rangle$ and
$\langle\phi_{N}|H|\phi_{N}\rangle$, respectively.

\begin{figure}
\includegraphics[width = 0.5\textwidth]{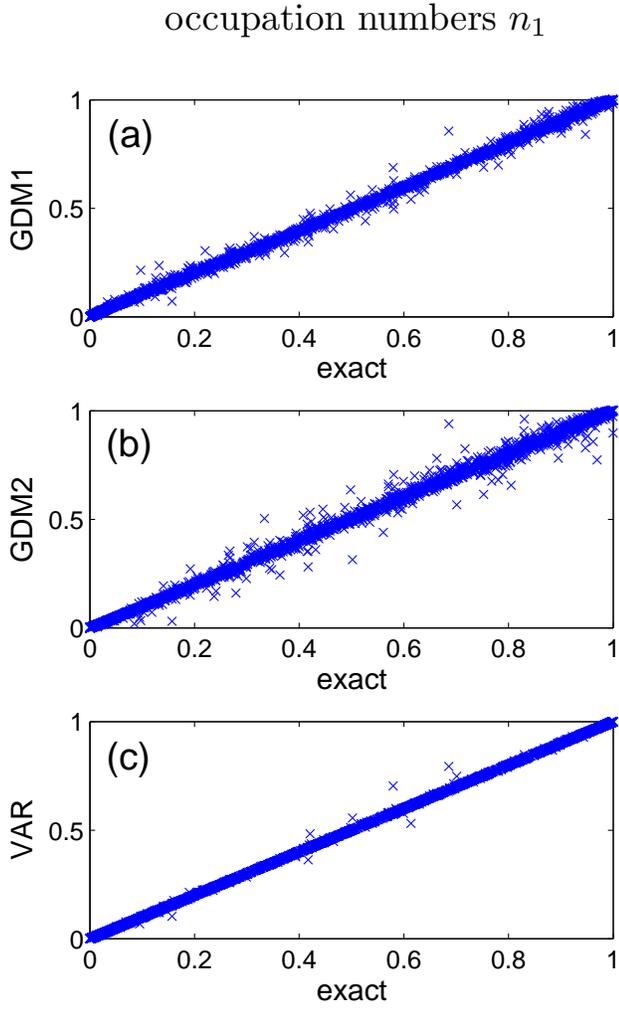}
\caption{\label{Fig_n1_line} (Color online) Occupation numbers $n_1$
of all the $3154$ single-particle levels in the ensemble by
different calculations. In panel (a), the single-particle levels and
the points are in one-to-one correspondence, with the horizontal
coordinate of the point being the exact $n_1$ and the vertical
coordinate being the GDM1 $n_1$. Similarly for panels (b) and (c),
but the vertical coordinates are $n_1$ of the GDM2 and the
variational calculations, respectively. See text for details.}
\end{figure}

\begin{figure*}
\includegraphics[width = 1.0\textwidth]{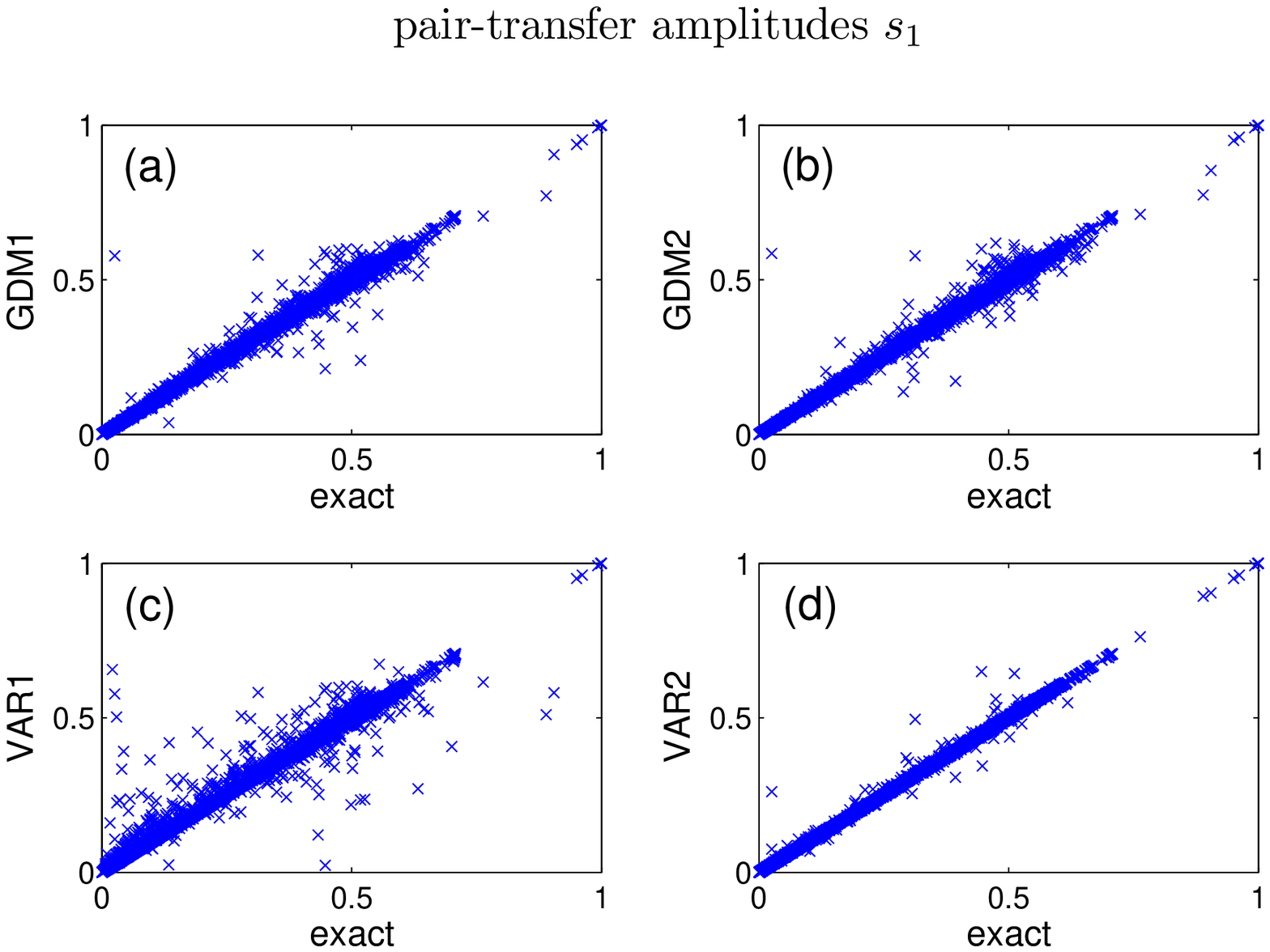}
\caption{\label{Fig_s1_line} (Color online) Pair-transfer amplitudes
$s_1$ of all the $3154$ single-particle levels in the ensemble by
different calculations. The points are plotted in the same way as
those in Fig. (\ref{Fig_n1_line}), but for the transfer amplitudes
$s_1$. Panels (c) and (d) plot two sets of variational calculations
(see text for details). }
\end{figure*}

In Figs. \ref{Fig_n1_line} and \ref{Fig_s1_line} I show the complete
spectroscopic results for the ensemble. For example, in panel (a) of
Fig. \ref{Fig_n1_line} there are $3154$ points (crosses),
corresponding to the $3154$ single-particle levels in the $1000$
examples of the ensemble. The horizontal coordinate of each point is
the exact value of $n_1$ of the corresponding single-particle level,
while the vertical coordinate is the ``GDM1'' value. Thus a perfect
calculation would have all the points lying on the $y = x$ straight
line. Similarly, in panels (b) and (c) of Fig. \ref{Fig_n1_line} the
vertical coordinates are the ``GDM2'' and ``VAR'' values of $n_1$,
respectively. Figure \ref{Fig_s1_line} is plotted in the same way
for the pair-transfer amplitudes $s_1$. From Figs. \ref{Fig_n1_line}
and \ref{Fig_s1_line} we see that the variational calculation for
$n_1$ and the ``VAR2'' version of $s_1$ are generally better than
the GDM ones. However, the less-careful ``VAR1'' calculation of
$s_1$ is worse than the GDM ones. The root-mean-square ($\sigma$)
derivations from the exact results are
\begin{eqnarray}
\sigma_n^{\rm{GDM1}} = 0.0122,~ \sigma_n^{\rm{GDM2}} = 0.0188,~
\sigma_n^{\rm{VAR}} = 0.0045,~  \label{rms_n}
\end{eqnarray}
on Fig. \ref{Fig_n1_line} and
\begin{eqnarray}
\sigma_s^{\rm{GDM1}} = 0.0207,~ \sigma_s^{\rm{GDM2}} = 0.0191,
\nonumber \\
\sigma_s^{\rm{VAR1}} = 0.0373,~ \sigma_s^{\rm{VAR2}} = 0.0091,
\label{rms_s}
\end{eqnarray}
on Fig. \ref{Fig_s1_line}. We note that on Fig. \ref{Fig_s1_line}
there is a point much worse than others for all the calculations, so
let us look at the particular example it belongs to. This example
has $2N = 12$ particles on two single-particle levels with angular
momenta $j = \frac{5}{2}$, $\frac{9}{2}$ and energies
$\epsilon_{\frac{5}{2}} = -8.988$, $\epsilon_{\frac{9}{2}} = -9.390$
MeV. The pairing two-body matrix elements are
$G_{\frac{5}{2},\frac{5}{2}} = 0.4252$, $G_{\frac{9}{2},\frac{9}{2}}
= 0.0456$, and $G_{\frac{5}{2},\frac{9}{2}} =
G_{\frac{9}{2},\frac{5}{2}} = 0.0016$ MeV. The failed point
corresponds to the $j = \frac{5}{2}$ level. This example is
particular in that $\epsilon_{\frac{5}{2}} - \epsilon_{\frac{9}{2}}
\approx G_{\frac{5}{2},\frac{5}{2}} \gg G_{\frac{9}{2},\frac{9}{2}}
\gg G_{\frac{5}{2},\frac{9}{2}}$, thus there is little correlation
between the two levels. I have looked at the exact wavefunction of
the daughter nucleus and it is mainly $P_{\frac{5}{2}}^\dagger
(P_{\frac{9}{2}}^\dagger)^4 |0\rangle$, which is not representable
by Eq. (\ref{gs}). Without this example, the root-mean-square
derivations for the pair-transfer amplitudes $s_1$ are
$\sigma_s^{\rm{GDM1}} = 0.0177$, $\sigma_s^{\rm{GDM2}} = 0.0163$,
$\sigma_s^{\rm{VAR1}} = 0.0352$, and $\sigma_s^{\rm{VAR2}} =
0.0079$.

The necessary formula for the GDM and variational calculations were
given in Ref. \cite{Jia_4}. In general, in large model spaces the
GDM calculation costs less time than the variational one, by a
factor of the number of non-degenerate single-particle levels (in
order of magnitude), because the former needs to calculate only
$\langle \phi_{N-1} | P_1 | \phi_N \rangle$ while the latter needs
$\langle \phi_N | P_1^\dagger P_2 | \phi_N \rangle$ ($P_1^\dagger
\equiv a_1^\dagger a_{\tilde{1}}^\dagger$). However, as we can see
from Figs. \ref{Fig_n1_line} and \ref{Fig_s1_line} and Eqs.
(\ref{rms_n}) and (\ref{rms_s}), the accuracy of the GDM method is
close to that of the variational one. The reduction of time cost may
be a big advantage when doing ab-initio mean-field calculations for
medium and heavy nuclei, especially if we were fitting parameters of
the interaction (for example the effort in developing density
functionals with spectroscopic accuracy).

\begin{figure}
\includegraphics[width = 0.5\textwidth]{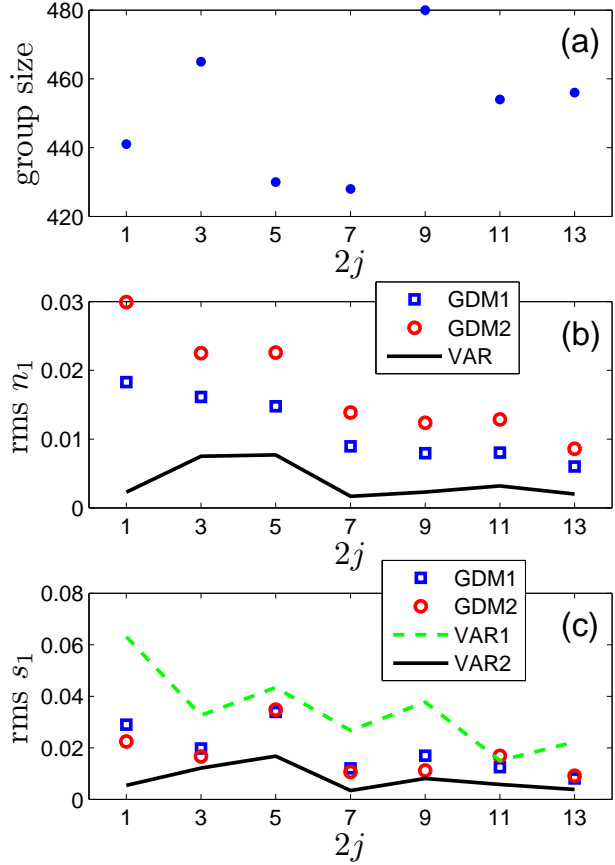}
\caption{\label{Fig_sub_2j} (Color online) Results of different
calculations grouped by the single-particle-level angular momentum
$j$. I divide all the $3154$ single-particle levels in the ensemble
into different groups according to their angular momentum $j$, and
panel (a) plots the group sizes. Panel (b) plots the
root-mean-square derivations from the exact results of the
occupation numbers $n_1$ by different calculations (GDM1, GDM2, VAR)
within each $j$ group. Similarly, panel (c) plots the
root-mean-square derivations of the pair-transfer amplitudes $s_1$
by four sets of calculations (GDM1, GDM2, VAR1, VAR2). See text for
details. }
\end{figure}

\begin{figure}
\includegraphics[width = 0.5\textwidth]{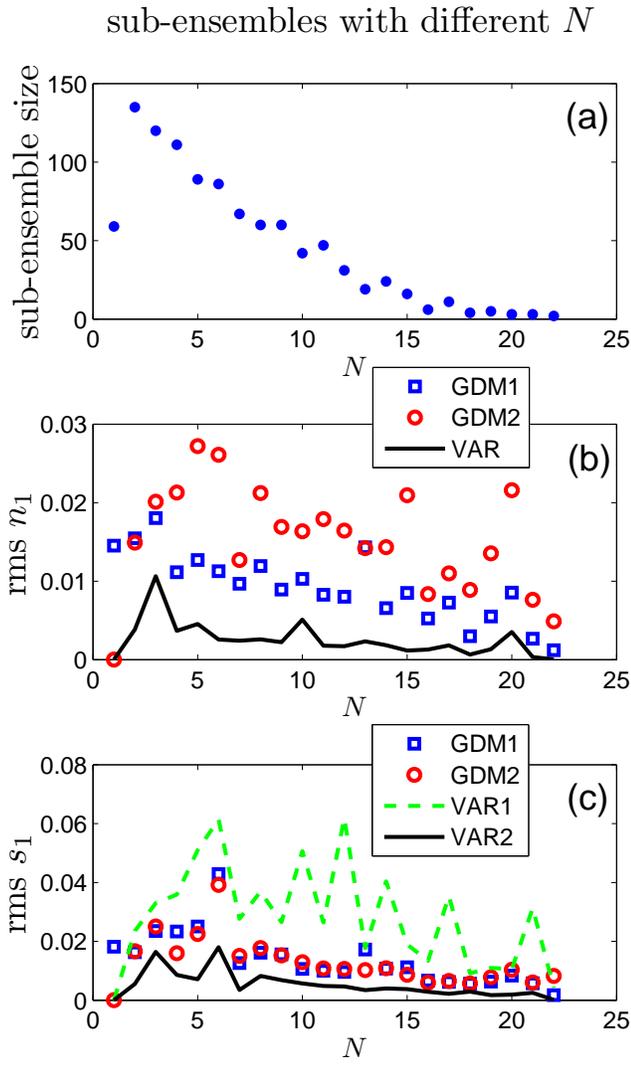}
\caption{\label{Fig_sub_N} (Color online) Results of different
calculations in sub-ensembles divided by the particle number $2N$.
The 1000 examples in the ensemble are divided into sub-ensembles
according to their particle number $2N$, and panel (a) plots the
sizes of the sub-ensembles. Panels (b) and (c) plot the
root-mean-square derivations by different calculations of the
occupation numbers $n_1$ and pair-transfer amplitudes $s_1$ within
each $N$ sub-ensemble. See text for details. }
\end{figure}

\begin{figure}
\includegraphics[width = 0.5\textwidth]{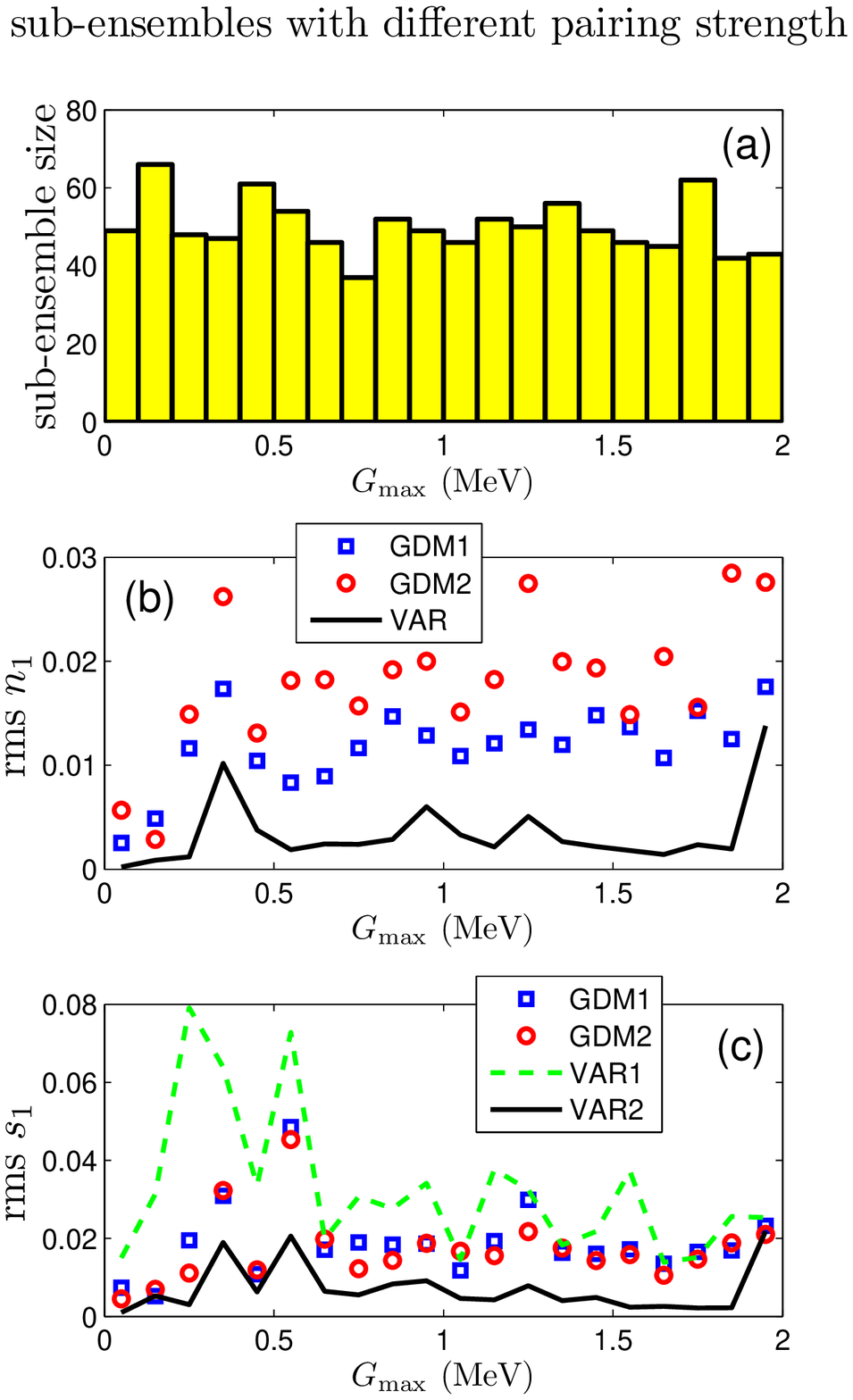}
\caption{\label{Fig_sub_G} (Color online) Results of different
calculations in sub-ensembles divided by the pairing strength
$G_{\rm{max}}$. The 1000 examples in the ensemble are divided into
$20$ sub-ensembles according to their pairing strength
$G_{\rm{max}}$, and panel (a) plots the sizes of the sub-ensembles.
Panels (b) and (c) plot the root-mean-square derivations by
different calculations of the occupation numbers $n_1$ and
pair-transfer amplitudes $s_1$ within each sub-ensemble. See text
for details. }
\end{figure}

Next I would like to see the accuracy of the methods depending on
different quantities. For this purpose I plot the root-mean-square
derivations of $n_1$ and $s_1$ as functions of the single-particle
angular momentum $j$, particle number $2N$, and pairing strength
$G_{\rm{max}}$ in Figs. \ref{Fig_sub_2j}, \ref{Fig_sub_N}, and
\ref{Fig_sub_G}, respectively. We have the following observations:
1. From Figs. \ref{Fig_sub_2j} and \ref{Fig_sub_N} we see that as a
trend the GDM results improve with increasing $j$ and $N$, which
should be expected because the GDM formalism is a collective theory.
2. There seems to be no obvious trend with the pairing strength
shown in Fig. \ref{Fig_sub_G}. 3. On Fig. \ref{Fig_sub_N}, the GDM2
calculation for $N = 1$, which is the particle-particle random phase
approximation, is much better than the GDM1 calculation. And for
small $N$ (until $N = 6$), the GDM2 $s_1$ seems to be slightly
better than the GDM1 $s_1$. 4. On Fig. \ref{Fig_sub_2j}, the GDM2
$s_1$ is slightly better than the GDM1 $s_1$ at the smallest angular
momenta $j = \frac{1}{2}$ and $\frac{3}{2}$. For these levels the
difference $n^N_1 - n^{N-1}_1$, which is inversely proportional to
$j$ around the Fermi surface, is largest; and the GDM2 theory with
$n^{N-1}_1$ seems to be slightly better than the GDM1 one with
$n^{N}_1$. 5. The GDM1 $n_1$ seems to be consistently better than
the GDM2 $n_1$ and the reason is still unclear. 6. In panel (c) of
Figs. \ref{Fig_sub_2j}, \ref{Fig_sub_N}, and \ref{Fig_sub_G} the
worst points are at $2j = 5$, $N = 6$, and $0.5$ MeV $< G_{\rm{max}}
< 0.6$ MeV, respectively; because these three sub-groups contain the
``worst'' example mentioned below Eq. (\ref{rms_s}).

\begin{figure}
\includegraphics[width = 0.5\textwidth]{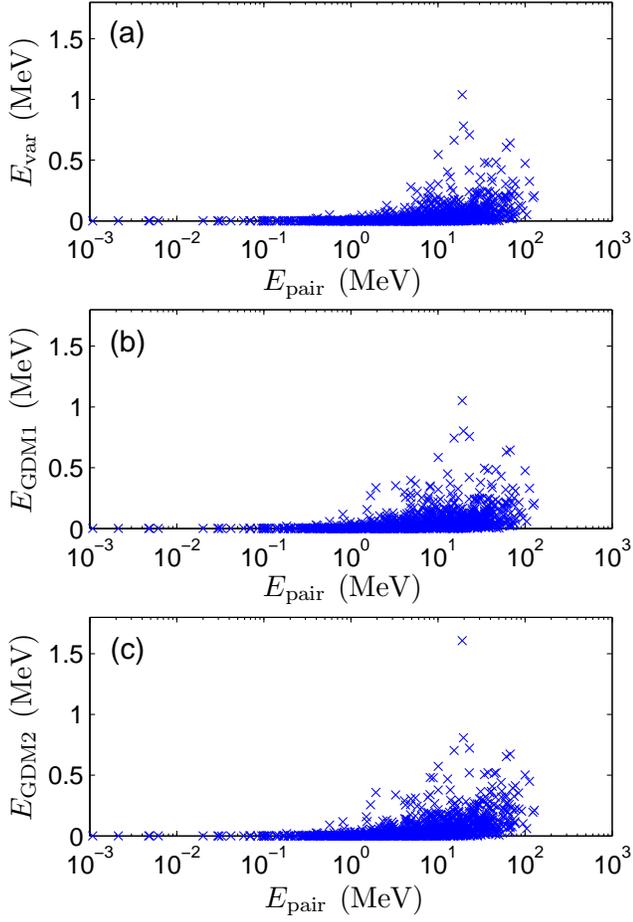}
\caption{\label{Fig_Egs} (Color online) Ground state energies of all
the $1000$ examples by different calculations. In panel (a), the
examples and the points are in one-to-one correspondence, with the
horizontal coordinate of the point being the pairing correlation
energy $E_{\rm{pair}}$, and the vertical coordinate being the
variational ground state energy $E_{\rm{var}}$ measured from the
exact ground state energy. Similarly for panels (b) and (c), but the
vertical coordinates are ground state energies by the GDM1 and GDM2
calculations, respectively, measured from the exact ground state
energy. See text for details. }
\end{figure}

At last in Fig. \ref{Fig_Egs} I show the results for the ground
state energy by different calculations. In panel (a) there are
$1000$ points (crosses), corresponding to the $1000$ examples of the
ensemble. The horizontal coordinate of each point is the pairing
correlation energy of the corresponding example, $E_{\rm{pair}}
\equiv \sum_{1} \epsilon_1 n^F_1 - E_{\rm{shell}}$, where
$E_{\rm{shell}}$ is the exact ground state energy of the shell model
calculation, and $n^F_1 = 1$ or $0$ is the occupation number of the
naive Fermi distribution. The vertical coordinate shows the
variational ground state energy measured from the exact one,
$E_{\rm{var}} = \langle \phi^{\rm{var}}_N | H | \phi^{\rm{var}}_N
\rangle - E_{\rm{shell}}$, where $| \phi^{\rm{var}}_N \rangle$ is
the pair condensate (\ref{gs}) with its pair structure $v_1$
(\ref{P_dag}) determined by the variation principle. Similarly for
panels (b) and (c), but the vertical coordinates are the ground
state energies of the GDM1 and GDM2 calculations, respectively,
measured from the exact one ($E_{\rm{GDM1}} = \langle
\phi^{\rm{GDM1}}_N | H | \phi^{\rm{GDM1}}_N \rangle -
E_{\rm{shell}}$ and $E_{\rm{GDM2}} = \langle \phi^{\rm{GDM2}}_N | H
| \phi^{\rm{GDM2}}_N \rangle - E_{\rm{shell}}$). We see that all the
three calculations give good ground state energies: the errors are
small relative to the pairing correlation energy. The average values
are $\bar{E}_{\rm{var}} = 0.045$, $\bar{E}_{\rm{GDM1}} = 0.061$,
$\bar{E}_{\rm{GDM2}} = 0.070$, and $\bar{E}_{\rm{pair}} = 13.64$
MeV. It is well known that the variational calculation finds the
best and lowest ground state energy for a set of restricted
wavefunctions of the form (\ref{gs}). However, from the above
average values we see that the energies of the GDM wavefunctions $|
\phi^{\rm{GDM1}}_N \rangle$ and $| \phi^{\rm{GDM2}}_N \rangle$ are
close to the variational minimum $\langle \phi^{\rm{var}}_N | H |
\phi^{\rm{var}}_N \rangle$.

\section{Summary  \label{Sec_sum}}

In summary, I derive a physically improved version of the GDM
pairing theory proposed in Ref. \cite{Jia_4}. Both versions are
checked in a large random ensemble, and the accuracy is given
statistically in terms of root-mean-square derivations from the
exact results. Consequently, we could consider the theories to be
correct and apply them with confidence to realistic systems.

Based on the results, the GDM theories are not as accurate as
(although close to) the variation principle. However, the reduction
of computing-time cost is huge for large model spaces (by a factor
of the number of non-degenerate single-particle levels). This should
be interesting for ab-initio mean-field calculations of medium and
heavy nuclei (especially if deformed Nilsson single-particle levels
were used), or the effort in fitting parameters of an interaction.

Comparing the two versions of the GDM theory, we see that in general
the new one (physically more reasonable) is slightly better in
calculating pair-transfer amplitudes $s_1$, while the old one
produces better occupation numbers $n_1$ and slightly better ground
state energy. The reason for the latter
is still unclear.

The key approximation of the current GDM methods is the
``factorization'' or ``linearization''of the two-body density matrix
on the pair condensate [Eqs. (\ref{fac1}) and (\ref{fac2})]. It
would be interesting to see its validity in other circumstances, in
particular, whether we could use it in the variational formalism
when calculating the two-body part of the average energy, which
would reduce the time cost to the same level of the current GDM
methods.\\

Support is acknowledged from the startup funding for new faculty
member in University of Shanghai for Science and Technology. The
numerical calculations of this work are done at the High Performance
Computing Center of Michigan State University.

%
%

%
%
%
%

\end{document}